\documentclass[pre,twocolumn,superscriptaddress,showpacs,showkeys,floatfix]{revtex4}
\usepackage{listings}
\usepackage{fancyvrb}
\usepackage{graphicx}
\usepackage{latexsym}
\usepackage{bm}
\usepackage{natbib}
\usepackage{dcolumn}
\usepackage{amsmath}
\usepackage{amstext}
\usepackage[english]{babel}
\usepackage{flafter}
\usepackage{url}
\usepackage{multirow}
\usepackage{float}
\newfloat{figtable}{t}{lop}
\floatname{figtable}{Table}
\newcommand{\Ql}{Q_{\lambda}}
\newcommand{\DQl }{\Delta Q_{\lambda}}
\newcommand{\ms }{\mathcal{S}}
\newcommand{\mc }{\mathcal{C}}

\begin{document}

\title{Limits of modularity maximization in community detection}

\author{Andrea Lancichinetti}
\affiliation{Complex Networks and Systems Lagrange Lab, Institute for
  Scientific Interchange, Torino, Italy}
\affiliation{Physics Department, Politecnico di Torino,
Corso Duca degli Abruzzi 24, 10129 Torino, Italy}
\author{Santo Fortunato}
\affiliation{Department of Biomedical Engineering and Computational
  Science, School of Science, Aalto University, P.O. Box 12200, FI-00076, Espoo,
  Finland}
\affiliation{Complex Networks and Systems Lagrange Lab, Institute for Scientific Interchange, Torino, Italy}
\begin{abstract} 
Modularity maximization is the most popular technique for the
detection of community structure in graphs. The resolution limit of
the method is supposedly solvable with the introduction of modified
versions of the measure, with tunable resolution parameters. We show
that multiresolution modularity suffers from two opposite coexisting problems: the
tendency to merge small subgraphs, which dominates when the resolution
is low; the tendency to split large subgraphs, which dominates
when the resolution is high. In benchmark networks with heterogeneous
distributions of cluster sizes, the simultaneous elimination of both
biases is not possible and multiresolution modularity is not capable
to recover the planted community structure, not even when it is
pronounced and easily detectable by other methods, for any value of the
resolution parameter. This holds for other multiresolution techniques
and it is likely to be a general problem of methods based on global optimization.
\end{abstract}



\maketitle

\section{Introduction}
\label{intro}

The detection and analysis of communities in graphs~\cite{girvan02, fortunato10} is one of the most
popular topics within the modern science of
networks~\cite{albert02,dorogovtsev01,newman03,pastor04,boccaletti06,caldarelli07,barrat08,cohen10}. In the latest years an
increasing number of large networked datasets including millions or
even billions of vertices and edges have become available, and a
traditional analysis based on local network properties 
and their global statistics (e.g., degree distributions and the like)
provides but a partial description of the system and its
function. Communities (also called {\it clusters} or {\it modules}) are subgraphs including vertices with similar
features or function, and their identification may disclose not only
such similarities among vertices, which are often hidden, but also how
the system is internally organized and works.

Vertices belonging to the same community have a considerably higher
probability of being linked to each other than vertices belonging  to
different clusters. Therefore a community appears as a region of the
network with a high density of internal links, much higher than the
average link density of the graph.
The most popular method to detect communities in graphs consists in
the optimization of a quality function, the modularity introduced by
Newman and Girvan~\cite{newman04b,newman06}. Modularity quantifies the
deviation of the internal link density of the clusters from the
density one expects to find within the same groups of vertices in
random graphs with the same expected degree sequence of the network at study.
The idea is that vertices linked to each other in a random way should
not form communities, as high values of the link density cannot be
attained. Consequently, high values of modularity are supposed to
indicate ``suspiciously''  high
values of internal link densities for the subgraphs, which are then
distinct from groups of randomly linked vertices and can be deemed as
true communities. While this is actually not
true~\cite{guimera04,reichardt06b}, the optimization of the measure
has been widely used in the past years.

Recently it has been pointed out that modularity optimization has a
number of problems. In particular, it has a
resolution limit~\cite{fortunato07}, that leads to the systematic
merger of small clusters in larger modules, even when the clusters are
well defined and loosely connected to each other. 
A more recent analysis of the resolution limit has led to the conclusion
that the modularity landscape is ``glassy'', and includes an
exponentially growing (with system size) number of local maxima whose
values are very close to the absolute maximum of the measure, even if
the corresponding partitions may be topologically quite different from
each other~\cite{good10}. This implies on the one hand that it is not
too difficult to find a good approximation of the modularity maximum
for many techniques, on the other hand that the maximum is essentially unreachable.
A recent comparative analysis of community finding algorithms has
indeed revealed that modularity fails to properly identify clusters on
benchmark graphs with built-in community structure, and that other
methods are much more effective~\cite{lancichinetti09c}.

Nevertheless, modularity optimization is still being used. The main
reason is the claim that the resolution limit can be
removed by adopting suitable multiresolution versions of modularity,
like those introduced by Reichardt and Bornholdt~\cite{reichardt06}
and by Arenas, Fern\'andez and G\'omez~\cite{arenas08b}. In these
variations, a tunable resolution parameter enables one to set the size of the
clusters to arbitrary values, from very large to very small. However,
real networks are characterized by the coexistence of clusters of very
different sizes, whose distributions are quite well described by power laws~\cite{clauset04,palla05,radicchi04}.
Therefore there is no characteristic cluster size and tuning a
resolution parameter may not help. Indeed, in this paper we show that
multiresolution modularity is not capable to identify the right
partition of the network in realistic settings and that therefore it
does not solve the problems of modularity maximization in practical
applications. The problem is that modularity maximization is not only
inclined to merge small clusters, but also to break large clusters,
and it seems basically impossible to avoid both biases
simultaneously. This applies to other multiresolution methods as well
and is probably a general feature of methods based on the optimization
of a global measure.

The paper is structured as follows. In Section~\ref{sec2} we present a
general analysis of some relevant mathematical properties of 
multiresolution modularity, with respect to
the merger or split of subgraphs, leading to the identification of a
range of values of the resolution parameter where modularity should be
safe from the above-mentioned problems. In Section~\ref{sec3} we test
the result on realistic benchmark graphs with community structure, showing
that it is often impossible to find a value of the resolution parameter that delivers the
planted partition. Conclusions are reported in Section~\ref{sec4}. 

\section{The problem of merging and splitting clusters}
\label{sec2}

\subsection{Multiresolution modularity}
\label{subsec2_1}

Our conclusions are not significantly affected by the
specific modularity formula one chooses, as we will show in
Section~\ref{sec3}. For the analytical discussion
of this Section we adopt
the generalized modularity $Q_{\lambda}$ proposed by Reichardt and
Bornholdt~\cite{reichardt06}, which reads 
\begin{equation}
\label{RB_modularity}
Q_{\lambda}=  \sum_{S} \left[\frac{k_{in}^S}{2M} - \lambda \Big(\frac{k_{tot}^S}{2M}\Big)^2\right],
\end{equation}
where the sum runs over all the clusters, $2M$ is the total degree of
the network, $k_{tot}^S$ is the sum of the degrees of vertices in
module $S$ and $k_{in}^S$ is twice the number of internal edges in
module $S$. So, we have $k_{tot}^S=k_{in}^S$ only if the module is
disconnected from the rest of the graph. Here 
$\lambda$ works like a resolution parameter: high values of $\lambda$
lead to smaller modules 
because the term $(k_{tot}^S/2M)^2$ in the sum of
Eq.~(\ref{RB_modularity}) becomes more important and its minimization,
induced by the maximization of $\Ql$, favors smaller clusters. 

We ask when it is proficuous for modularity to keep two subgraphs together
or separate. For this, we 
need to compute the difference $\Delta \Ql = \Ql
(\textrm{partition with merged subgraphs}) 
- \Ql (\textrm{partition with separated subgraphs})$: if $\DQl>0$
modularity would be higher for the partition where the subgraphs are merged, otherwise the split
would be more convenient. 
\begin{figure}
\begin{center}
\includegraphics[width=\columnwidth]{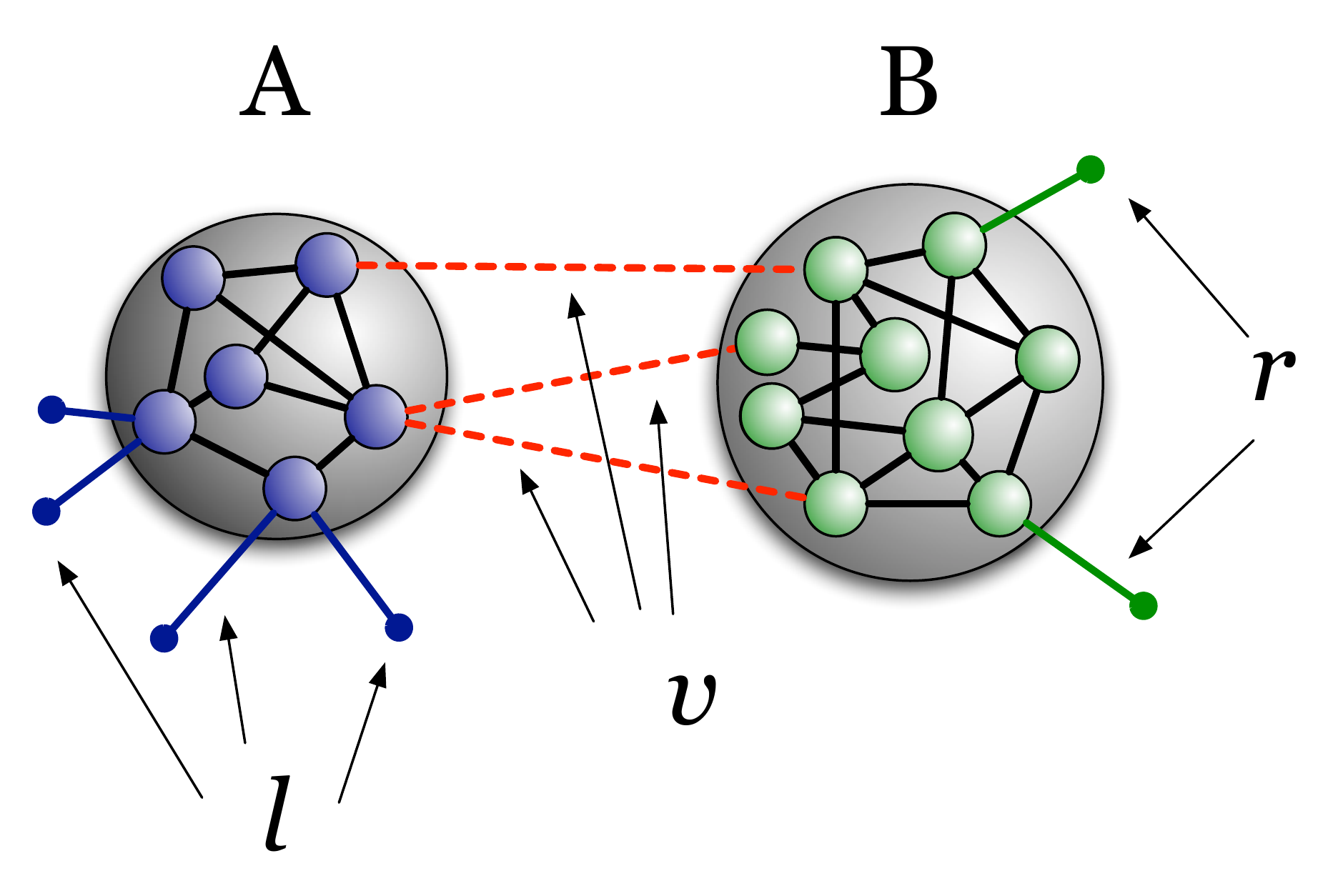}
\caption{(Color online) Schematic representation of the problem of merging versus
splitting subgraphs. Here $A$ and $B$ are two subgraphs, the problem
is whether one yields a higher value for modularity by merging them
in a single subgraph or by keeping them separated. The parameters
involved in the decision are the number of internal links in $A$ and
$B$ (multiplied by $2$), $k_{in}^A$ and
$k_{in}^B$, the number of links $v$ between $A$ and $B$ (here $v=3$),
the number of links $l$ between $A$ and vertices
belonging neither to $A$ nor to $B$ (here $l=4$), and its equivalent $r$
for $B$ (here $r=2$). }
\label{fig1}
\end{center}
\end{figure}

We indicate with $A$ and $B$ the two subgraphs (see
Fig.~\ref{fig1}). Let $\Ql^{A-B}$ and $\Ql^{AUB}$ denote the value of
modularity when $A$ and $B$ are kept separated and merged, respectively.
\begin{eqnarray}
\label{qsep}
\Ql^{A-B}&=& \Big[\sum_{S \neq A, B} \dots \Big]+\frac{k_{in}^A}{2M} +
\frac{k_{in}^B}{2M} \nonumber\\
&-& \lambda \Big(\frac{k_{in}^A + l + v}{2M}\Big)^2  -  \lambda \Big(\frac {k_{in}^B + r + v}{2M}\Big)^2, 
\end{eqnarray}
where $v$ denotes  the number of links joining $A$ with $B$, $l$ the
number of links joining $A$ with the rest of the network (excluding $B$)
and $r$ is the 
equivalent of $l$ for $B$. For $\Ql^{AUB}$ we have:
\begin{eqnarray}
\label{qtog}
\Ql^{AUB}&=& \Big[\sum_{S \neq A, B} \dots
\Big]+\frac{k_{in}^A}{2M} +  
\frac{k_{in}^B}{2M} + \frac{2 v}{2M} \nonumber\\
&-& \lambda \Big(\frac{k_{in}^A + l + v + k_{in}^B +r + v}{2M}\Big)^2.
\end{eqnarray}
The difference $\DQl= \Ql^{AUB} - \Ql^{A-B}$ reads
\begin{eqnarray}
\label{qdelta}
\DQl&=&  \frac{2 v}{2M} - \lambda \frac{k_{in}^A k_{in}^B  + l
  k_{in}^B + r k_{in}^A  + l r}{2M^2} \nonumber\\
&-&\lambda \frac{v (k_{in}^A + k_{in}^B + l + r   ) + v^2 }{2M^2}.
\end{eqnarray}
To simplify a little Eq.~(\ref{qdelta}) we can define $\Delta= 2M \DQl$
\begin{eqnarray}
\label{delta_eq_1}
\Delta&=&  2 v - \lambda \frac{k_{in}^A k_{in}^B  + l k_{in}^B + r
  k_{in}^A  + l r}{M} \nonumber\\
&-&\lambda\frac{v (k_{in}^A + k_{in}^B + l + r   ) + v^2 }{M}.
\end{eqnarray}
Modularity is higher for $A$ and $B$ merged if and only if $\Delta >0$.

Eq.~(\ref{delta_eq_1}) is rather general but we are just interested in
testing modularity for some special cases, for which calculations are easy. 
Here in particular, we will consider the case $l=r=\eta$ and
$k_{in}^A=k_{in}^B=\xi$.  
Eq.~(\ref{delta_eq_1}) becomes
\begin{equation}
\label{delta_eq}
\Delta=  2 v - \lambda \frac{(\xi+v +\eta)^2 }{M}.
\end{equation}

These results are essential to follow the discussion of the next subsections.

\subsection{Splitting clusters}
\label{subsec2_2}

Despite the different approaches to the problem of detecting clusters
in networks, there are some general ideas which are shared by most
scholars. One of them is that a random graph has no
communities, so it should not be split by an algorithm in smaller
pieces, with the only exception of the trivial split in singletons,
i.e. in groups containing each just a single vertex, which is still an acceptable
answer. 
Another shared belief is that a complete graph (or clique), i.e. a graph whose
vertices are all connected to each other, is a perfect community (due
to the fact that the internal link density reaches the highest
possible value of $1$). So, if cliques are just loosely
connected to each other, one would expect that a good method should
detect them as separate clusters. We would like to find the
mathematical conditions, in particular the choice of the resolution
parameter $\lambda$, that satisfy both requisites. In this subsection
we search for the condition to avoid the splitting of random
subgraphs, while the condition to avoid the merger of cliques will be
given in the next subsection.

Let us consider a random subgraph
$\ms$ with total degree $2M_\ms$, which is part of a larger
network with total degree $2M$. The goal is to check under which condition 
$\ms$ is split by optimizing modularity. Here for simplicity we
consider only bi-partitions. The expected optimal modularity $Q_2$ for the
bipartition of a random graph has been computed by 
Reichardt and Bornholdt~\cite{reichardt07}
\begin{equation}
\label{RB_eq}
Q_{RB}= 0.765 \frac{\langle \sqrt{k} \rangle_\ms} {\langle k
  \rangle_\ms},
\end{equation}
where the brackets indicate expectation values over the ensemble of
random graphs with the same expected degree sequence of the subgraph
at study.

We now express $Q_2$ in terms of the number of edges $v$ between the
clusters of the bipartition with optimal modularity. We obtain
\begin{equation}
2 M_{\ms} Q_{2}= 2 M_{\ms} -2 v - \frac{k_A^2 + k_B^2}{2 M_{\ms} } =  \frac{2 k_A  k_B}{2 M_{\ms} } - 2 v,
\end{equation}
where $k_A$ ($k_B$) is the total degree of module $A$ ($B$). Since
modularity is optimal when the two modules are of about equal size,
i.e. when $k_A \approx k_B \approx M_\ms$, we have:
\begin{equation}
2 M_{\ms} Q_{2}= M_{\ms} - 2 v,
\end{equation}
from which we can derive $v$,
\begin{equation}
v = M_{\ms} \Big(  \frac{1}{2} -  Q_{2}\Big).
\label{eqv}
\end{equation}
\begin{figure}[h!]
\begin{center}
\includegraphics[width=\columnwidth]{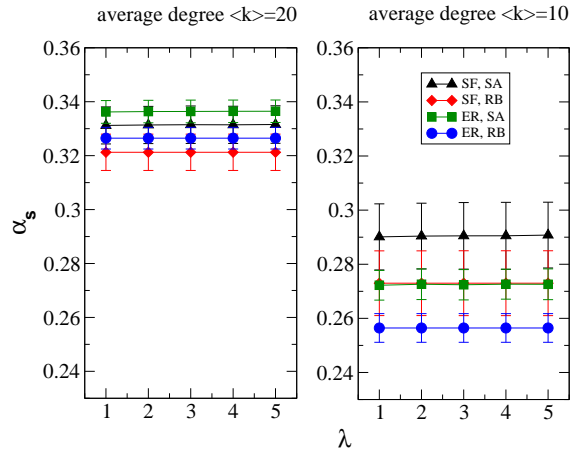}
\caption{(Color online) The plot shows $\alpha_\ms$ measured on Erd\"os-R\'enyi and
scale free graphs. For each type of graph we plot the analytical
estimate of Reichardt and Bornholdt (RB) and a numerical estimate
obtained by optimizing modularity with simulated annealing (SA)~\cite{guimera04}. The minimum cut $v=\alpha_\ms \times M_\ms$ was
measured by optimizing
modularity for different values of $\lambda$ over the set of bipartitions. To optimize modularity, we are looking for small 
values of $v$ and equal values of $k^A$ and $k^B$, so tuning $\lambda$
just controls the importance of either requirement. However,
simulations show that the dependence on $\lambda$ is quite weak, validating our approximation $k^A \approx k^B$.}
\label{FIG2}
\end{center}
\end{figure}
For $Q_{2}=0$ we would have $v = M_{\ms}/2$, which is the
expected average number of links joining two modules of equal size,
arbitrarily chosen. Eq.~(\ref{eqv}) implies that 
optimizing modularity decreases the number of expected links between
the modules, with respect to arbitrary bipartitions, while it
increases the internal density of links of the modules. One also
sees that, for $v$ to be positive, $Q_{2} \leq 0.5$. Actually, in the
calculation of Reichardt and Bornholdt, this holds only if $\langle k
\rangle$ is big enough. To give an idea of the
numbers that one could have, $Q_{2} \approx 0.17$ when all vertices have degree $k=20$, 
so $v \approx 0.33 \times M_\ms $ which is actually a not too bad approximation 
also for other degree distributions (for all vertices having degree $k=10$, $ v \approx 0.25
\times M_\ms $). Let us call $\alpha_{\ms}$ this proportionality factor between $v$
and $M_\ms$,
\begin{equation}
\label{alpha_defs}
 v=\alpha_{\ms} M_{\ms} \;\; \textrm{and} \;\;\; k_{in}^A=k_{in}^B=(1-\alpha_{\ms}) M_\ms.
\end{equation}
From Eqs.~(\ref{RB_eq}), (\ref{eqv}) and (\ref{alpha_defs}) we get
\begin{equation}
\label{alpha_defs_rb}
 \alpha_{\ms} = \frac{1}{2} - 0.765 \frac{\langle \sqrt{k} \rangle_\ms} {\langle k \rangle_\ms}.
\end{equation}
In Fig.~\ref{FIG2} we compare the values of $\alpha_{\ms}$ from
Eq.~(\ref{alpha_defs_rb}) with numerical estimates
derived by putting in Eq.~(\ref{eqv}) the maximum modularity $Q_2$,
derived with simulated annealing. The calculation of $Q_2$ is carried
out for different values of $\lambda$, but the results seem to be
essentially independent of $\lambda$. We consider both Erd\"os-R\'enyi (ER)
and scale-free (SF) graphs, with $1000$ vertices and average degree $\langle k\rangle=20$ (left
panel) and $10$ (right panel). The SF graphs have degree
exponent $2$. As we can see from Fig.~\ref{FIG2}, the analytical
estimate of Eq.~(\ref{alpha_defs_rb}) yields a good approximation of $\alpha_{\ms}$.

Let us now consider our splitting-merging problem, considering $A$ and
$B$ as candidates. We set $\eta=1$, which means that only two links come out of $\ms$ 
(ideally one from $A$, the other from $B$).
In this case, we would like to have $\Delta>0$, to avoid the split of
the random subgraph $\ms$. From
Eq.~(\ref{delta_eq}) and 
Eqs.~(\ref{alpha_defs}) we get (remember that $\xi=k_{in}^A=k_{in}^B$):
\begin{equation}
2 \alpha_{\ms} M_{\ms}> \frac{\lambda (M_\ms +1)^2}{M},
\end{equation}
which implies
\begin{equation}
\label{cond1}
\lambda < \frac{ 2 \alpha_{\ms} M}{M_\ms}.
\end{equation}
Alternatively, we can incorporate the correction factor
$[M_\ms/(M_\ms+1)]^2 \approx 1$ 
in $\alpha_\ms$, so that we call $\alpha_\ms$ what is actually $\alpha_\ms [M_\ms/(M_\ms+1)]^2$.
If the subgraph is a clique, $\alpha_\ms \approx 0.5$, and modularity can even split a clique when
\begin{equation}
\lambda \gtrsim \frac{M}{M_\ms}. 
\end{equation}

\subsection{Merging clusters}
\label{subsec2_3}

Let us now consider two equal sized subgraphs connected with one edge
($v=1$ and $\eta=1$) and let $k_{in}^A=k_{in}^B=\xi_\mc$.
Eq.~(\ref{delta_eq}) becomes:
\begin{equation}
\Delta=  2  - \lambda \frac{(\xi_\mc+2)^2}{M}.
\end{equation}
In this case we want $\Delta<0$ (we wish to keep the two subgraphs separated), which implies
\begin{equation}
\label{cond2}
\lambda  > \lambda_\mc=\frac{2M} {(\xi_\mc+2)^2}.
\end{equation}
If $\xi_\mc$ is very small, $\lambda$ has to be very big (for
$\lambda_\mc >1$ the subgraphs cannot be resolved by standard
modularity,  which corresponds to $\lambda=1$, and we recover the resolution limit of Ref.~\cite{fortunato07}). 
On the other hand if $\xi_\mc$ is large, the subgraphs will be resolved
for a large range of $\lambda$-values.

If the subgraphs are two cliques of $n_\mc$ nodes each, for instance, $\xi_\mc=n_\mc (n_\mc-1)$.

\subsection{Condition on the ineliminability of the bias}
\label{subsec2_4}

We now put together conditions (\ref{cond1}) and (\ref{cond2}).
We have that
\begin{equation}
\label{cond_together}
\lambda_2 < \lambda  < \lambda_1,
\end{equation}
where
\begin{equation}
\lambda_{1} =  \frac{2\alpha_\ms M}{M_\ms} \;\;\textrm{and}\;\;   \lambda_{2} = \frac{2M} {(\xi_\mc+2)^2}.
\end{equation}
Above $\lambda_{1}$, modularity splits random subgraphs, below
$\lambda_{2}$ it puts together subgraphs 
even if they are connected by just one link (even in the case in which
they are cliques). In the range between $\lambda_1$ and $\lambda_2$ it
should be possible to avoid both biases. However, if
\begin{equation}
\lambda_1 < \lambda_2, 
\label{lcondition}
\end{equation}
the biases cannot be both simultaneously lifted.
Eq.~(\ref{lcondition}) holds when, by setting $M_\ms/\alpha_{\ms}= \beta_\ms$,
\begin{equation}
\label{incomp}
{(\xi_\mc+2)^2} < \beta_\ms.
\end{equation}
Note that Eq.~(\ref{incomp}) does not depend on the size of the whole network, either
in terms of vertices or edges.

\begin{figure}[h!]
\begin{center}
\includegraphics[width=\columnwidth]{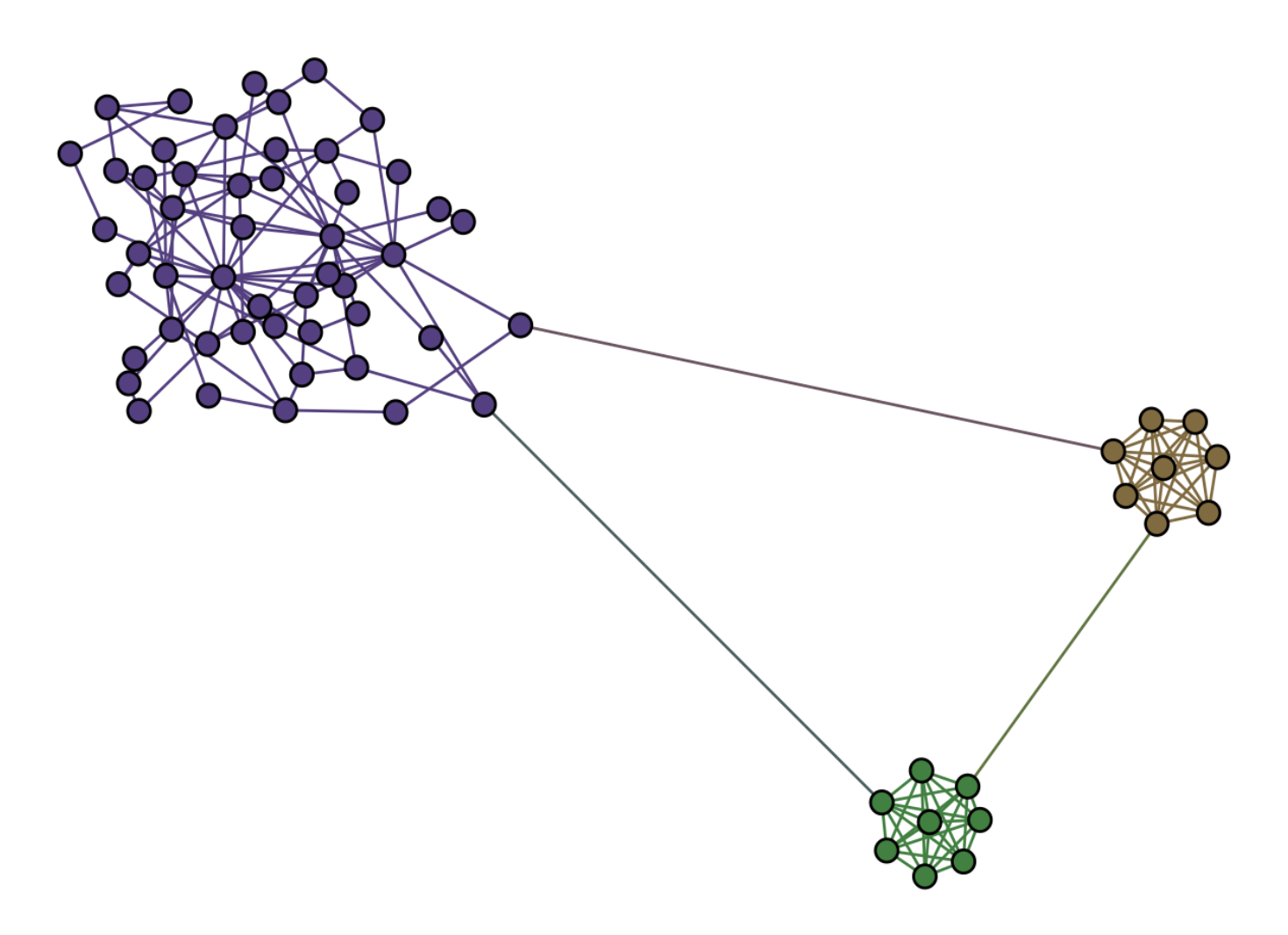}
\caption{(Color online) Schematic network with two cliques and a
  random subgraph, which are the natural communities of the network.}
\label{scheme_3}
\end{center}
\end{figure}

To be more concrete we consider a simple example. We examine a network
made out of two identical cliques of $n_\mc$ vertices each and an internally random
subgraph of $n_\ms$ vertices 
and average degree $\langle k \rangle _\ms$. The three clusters 
are all connected to each other by one edge only (see Fig.~\ref{scheme_3}). 
\begin{figure}[h!]
\begin{center}
\includegraphics[width=\columnwidth]{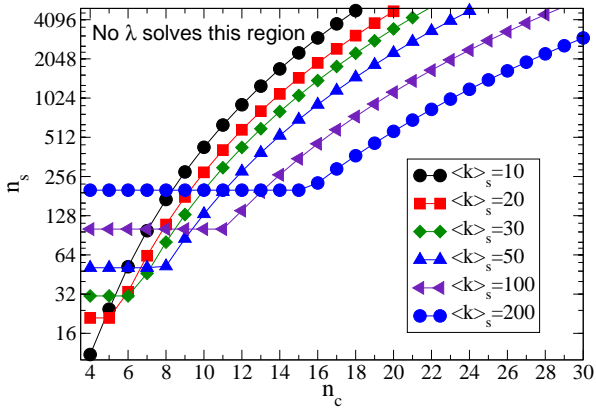}
\caption{(Color online) This plot shows Eq.~(\ref{incomp}) as a function of $n_\ms$
and $n_\mc$ for the simple network with three clusters of Fig.~\ref{scheme_3}
Above the curves modularity cannot find the right partition for any value of $\lambda$.}
\label{incomp_fig}
\end{center}
\end{figure}
In Fig.~\ref{incomp_fig} we
plot the relation between $n_\mc$ and $n_\ms$ coming from the
equality $\lambda_1=\lambda_2$ (obtained turning the inequality of
Eq.~(\ref{incomp}) to an equality) 
for some values of $\langle k \rangle _\ms$. We used Eq.~(\ref{alpha_defs_rb}) to
evaluate $\alpha_\ms$, with the approximation $\langle \sqrt{k}
\rangle_\ms =\sqrt{\langle k\rangle_\ms}$ 
and the relations $\xi_\mc=n_\mc (n_\mc-1)$ and $M_\ms=n_\ms \langle k
\rangle _\ms /2 $.
For any given value of $\langle k \rangle _\ms$, the inequality of Eq.~(\ref{incomp}) holds
above the corresponding curve.

\begin{figure}[h!]
\begin{center}
\includegraphics[width=\columnwidth]{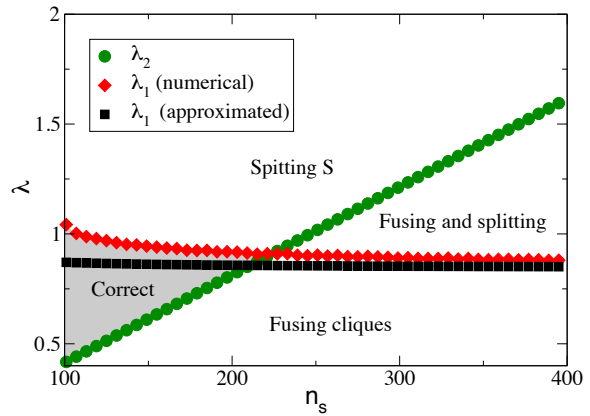}
\caption{(Color online) Threshold parameters $\lambda_1$ and
$\lambda_2$ as a function of $n_\ms$ ($n_\mc=13$, $\langle k\rangle_\ms=100$). The theoretical line for
$\lambda_1$ is obtained by approximating $\alpha_\ms$ as described in the text.
We see that $\lambda_1> \lambda_2$, up to $n_\ms \approx 230$, 
so that no $\lambda$ can eliminate the biases for bigger values of
$n_\ms$. When $n_\ms<\approx 230$, the biases can be both eliminated
only in the shadowed area between the curves.}
\label{colorplot2}
\end{center}
\end{figure}

In Fig.~\ref{colorplot2} we plot $\lambda_1$ and $\lambda_2$ as a
function of $n_\ms$, for $n_\mc=13$ and $\langle k\rangle_\ms=100$. For $\lambda_1$
we show two curves, one corresponding to the exact function, determined
numerically, while for the other we have used the theoretical approximation of
$\alpha_\ms$ described above. The lines divide
the $\lambda-n_\ms$ plane in four areas, characterized by the presence or absence of
the two biases. As we can see, the portion of the plane in which both
biases are simultaneously absent (gray area) is quite small.

One might still wonder that it could be possible to find a value of $\lambda$ high
enough that the 
random subgraph $\mathcal{S}$ is split in $n_\mathcal{S}$ vertices and the two cliques are still correctly detected. 
Let us consider Eq. $5$ when $A$ consists of a single vertex, so that
$v$ is the internal degree of the vertex with respect to $B$ and
$l+v=k^A$ is the total degree of $A$. 
Recalling that $k_{in}^B + r +v=k_{tot}^B$, Eq. $5$ becomes:
\begin{equation}
\Delta = 2 v   - \lambda \frac{k_{tot}^B k^{A}}{M}.
\end{equation}
Therefore $A$ and $B$ would be kept separated when:
\begin{equation}
\lambda > \frac{2M v}{k_{tot}^B k^{A}  }.
\end{equation}
By increasing $\lambda$ we can actually separate some vertices of
$\mathcal{S}$ and we would eventually split it in $n_\mathcal{S}$ clusters when
$\lambda > \frac{2M}{x}$, where $x$ is the minimum $k_i k_j$ over all the connected vertices
$(i, j)$ of $\mathcal{S}$.  Similarly, the condition for the cliques not to be split reads:
\begin{equation}
\lambda < \frac{2M }{(n_{\mathcal{C}}-1)  (n_{\mathcal{C}}-2)},
\end{equation}
since the denominator is the total degree of a clique of
$n_{\mathcal{C}}-1$ vertices (we neglected $r$) and 
we considered $k^A=v$ (the vertex does not have external connections).

In conclusion, if there are two connected vertices in $\mathcal{S}$
such that the product of their degrees is smaller than $(n_{\mathcal{C}}-1)(n_{\mathcal{C}}-2)$, no
values of $\lambda$ are suitable to guess the right answer(s). This is very likely 
to happen if the degree distribution of $\mathcal{S}$ is broad, so that there are many low-degree vertices.

\section{Tests on benchmark graphs}
\label{sec3}

We want now to check the practical consequences of the limits of
multiresolution modularity. For that we take the LFR benchmark, a model of graphs with built-in community structure 
that we have recently introduced~\cite{lancichinetti08}.
It is an extension of the {\it planted $\ell$-partition
  model} introduced by Condon and Karp~\cite{condon01}. Each
graph has power law distributions of degree and community size,
which are common features of real graphs with community structure.
The degree of mixture between clusters is measured by the {\it mixing
  parameter} $\mu$, expressing the ratio between the number of
neighbors of a vertex outside its community and the total number of
neighbors. So $\mu=0$ indicates that clusters are topologically
disconnected from each other, as each vertex has neighbors within its
community only, while $\mu=1$ indicates that vertices are connected
only to vertices outside their group, so the groups are not communities.
Vertices are linked to each other at random, compatible with the
constraints on the distributions of degree and community size and to
the fact that $\mu$ has to be (approximately) the same for all
vertices. So the clusters are essentially random subgraphs.

\begin{figure}[h!]
\begin{center}
\includegraphics[width=\columnwidth]{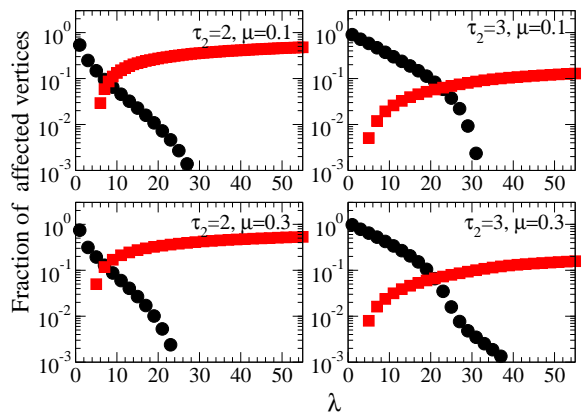}
\caption{(Color online) Test of multiresolution modularity on LFR
  benchmark graphs. Each panel shows the fraction of misclassified
  vertices due to artificial mergers (circles) and splits (squares) of
  clusters, as a function of the resolution parameter $\lambda$. The
  panels correspond to different choices of the exponent $\tau_2$ of the
  cluster size distribution of the graph and of the mixing parameter
  $\mu$. Each point represents an average over $100$ benchmark
  graphs. All graphs have $10000$ vertices. The other parameters are:
  average degree $\langle k\rangle =20$; maximum degree $k_{max}=100$;
  minimum cluster size $c_{min}=10$; maximum cluster size
  $c_{max}=1000$; degree exponent $\tau_1=2$.}
\label{bench1}
\end{center}
\end{figure}

\begin{figure}[h!]
\begin{center}
\includegraphics[width=\columnwidth]{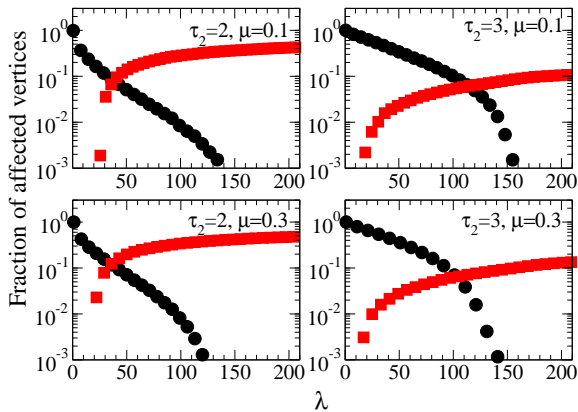}
\caption{(Color online) Same as Fig.~\ref{bench1}, but for LFR
  benchmark graphs of $50000$ vertices. All other parameters are the
  same as for the graphs used in Fig.~\ref{bench1}.}
\label{bench2}
\end{center}
\end{figure}

We want to specialize Eq.~(\ref{delta_eq_1}) to the LFR benchmark graphs.
Let us consider a cluster $\ms$ with $n_b$ nodes, total degree $2m_b$ 
and internal degree $2M_{\ms b}$. 
We split it into two equal-sized subgraphs such that the internal
degree of either part is the same: $k_{in}^A=k_{in}^B$. Moreover, for
simplicity we assume that the split is done such to keep an equal
number of edges between each of the subgraphs and the rest of the network: $l=r$.
We have $M_{\ms b}= (1-\mu)  m_b$, $l=r=\mu m_b$, $v= \alpha_{\ms b}
M_{\ms b} = \alpha_{\ms b} (1-\mu)  m_b$. 
The condition of non-splitting is:
\begin{equation}
2 v > \lambda \frac{(M_{\ms b}+l)^2 }{M},
\end{equation}
which is:
\begin{equation}
2  \alpha_{\ms b} (1-\mu)  m_b > \lambda \frac{m_b^2}{M}.
\end{equation}
So, 
\begin{equation}
 \lambda < \lambda_1  \;\;\; \textrm{where} \;\;\; \lambda_1=  2  \alpha_{\ms b}(1-\mu)\frac{M}{m_b }. 
\end{equation}
We now search for the condition that leads to the merger of 
two clusters of an LFR benchmark graph. For that we should know how many
edges they share, which depends on the graph size and the number of
clusters. We call $v_{xy}$ the number of edges between modules $x$ and
$y$ and $2m_x$ and $2m_y$ their total degrees. Eq.~(\ref{delta_eq_1}) becomes
\begin{equation}
\Delta=  2 v_{xy} - \lambda \frac{4 m_x  m_y}{M}.
\end{equation}
The condition to keep the clusters separated is $\lambda>\lambda_2$, where
\begin{equation}
\lambda_2=  \frac {M v_{xy}}{2 m_x m_y}. 
\end{equation}
So, the two biases can be simultaneously removed iff $\lambda_1 >
\lambda_2$, which amounts to
\begin{equation}
2  \alpha_{\ms b} (1-\mu)\frac{M}{ m_b }  > \frac {M v_{xy}}{2m_x
  m_y}. 
\label{finalineq}
\end{equation}
The inequality of Eq.~(\ref{finalineq}) has to hold for all triples of clusters
$x$, $y$ and $b$, and this is usually unlikely to happen. 
In order to show that, we check whether multiresolution modularity is
able to deliver the planted partition of the LFR benchmark graphs for any
value of the resolution parameter $\lambda$. The results are shown in
Figs.~\ref{bench1} and \ref{bench2}. We
plot the fraction of vertices which are incorrectly classified by
modularity as a function of $\lambda$. We just consider
misclassifications caused by merging (circles) or splitting (squares) the clusters of the
planted partition of the graphs. We see that, for small values of
$\lambda$, modularity merges many clusters and essentially splits none, whereas for large
$\lambda$ there is a dominance of splitting over merging. The plots
clearly show that, for every value of $\lambda$, there will be some
misclassification due to cluster merging, splitting or both. The
fraction of affected vertices does not go below $10\%$ but it can be considerably
larger. Fig.~\ref{bench1} refers to graphs with $10000$ vertices, but
the situation does not improve if we go to larger graph sizes ($50000$
vertices for the benchmark graphs used for Fig.~\ref{bench2}). We
point out that we have chosen low values of the mixing parameter
$\mu$ ($0.1$ and $0.3$), corresponding to clusters which are well
separated from each other. Modern
algorithms for community detection (like Infomap~\cite{rosvall08} and
OSLOM~\cite{lancichinetti11}) would easily find the correct
partitions in the graphs we have used for the tests of
Figs.~\ref{bench1} and \ref{bench2} (see
Ref.~\cite{lancichinetti09c}). One may object that our estimate
of the modularity maximum for each graph is just an approximation of the actual
result, whose search is an NP-complete
problem~\cite{brandes08}. However, we have checked in each case that
the partitions found have a higher modularity than the planted partition
of the benchmark graphs.

\begin{figure}[h!]
\begin{center}
\includegraphics[width=\columnwidth]{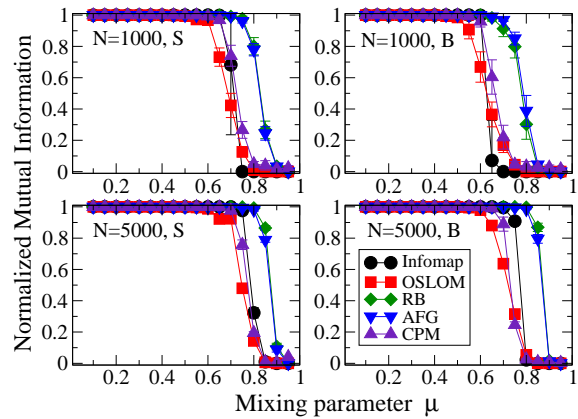}
\caption{(Color online) Comparative analysis of several
  multiresolution techniques on the LFR benchmark. The graphs are made
  of $1000$ and $5000$
vertices, the exponent of the degree distribution $\tau_1=2$, the
exponent of the
clusters size distribution $\tau_2=1$, the average
degree $\langle k\rangle=20$, the maximum degree $k_{max}=50$, the
cluster size ranges are $S=[10,
50]$ and $B=[20:100]$.}
\label{fig8}
\end{center}
\end{figure}

Finally we would like to check how general our results are. We have
focused on the multiresolution modularity proposed by Reichardt and
Bornholdt in Ref.~\cite{reichardt06}. In this paper, however, the
authors had proposed a general ansatz for the quality function, and
their multiresolution modularity was just a specific case of it. In a
recent work~\cite{traag11}, Traag et al. have shown that this ansatz
can be specialized to include other known measures, like the multiresolution
modularity by Arenas et al.~\cite{arenas08b}, and the quality function
adopted by Ronhovde and Nussinov~\cite{ronhovde10}, which is
characterized by not having a null model term, in contrast to
modularity. In fact, Traag et al. derived another model
from the general class of functions of Reichardt and Bornholdt, which
they called Constant Potts Model (CPM), which allegedly has no
resolution limit. In Fig.~\ref{fig8} we reproduce the results of the
comparative analysis performed by Traag et al. on the LFR benchmark.
Here we compare five methods: Infomap, OSLOM, the optimization of the
multiresolution modularities of Reichardt and Bornholdt (RB) and
Arenas et al. (AFG), and the CPM by Traag et al.. 
For each selected value of the mixing parameter $\mu$ we generated
$100$ realizations of the LFR benchmark, and averaged on them the values of
the similarity between the detected and the planted partition. As similarity measure we took the Normalized Mutual
Information (NMI)~\cite{danon05}, which has become a standard in this kind
of evaluations. In our computations we used a modified version of the
measure~\cite{lancichinetti09}, recently introduced by the authors of this paper, that is
able to estimate the similarity of partitions as well as the
similarity of covers, i.e., of divisions of a
network into overlapping communities. We have used this version of the
NMI in our comparative analysis of community detection
algorithms~\cite{lancichinetti09c}, so we stick to it for
consistency. We stress however that the clusters of the graphs we
considered are not overlapping.

\begin{figure}[h!]
\begin{center}
\includegraphics[width=\columnwidth]{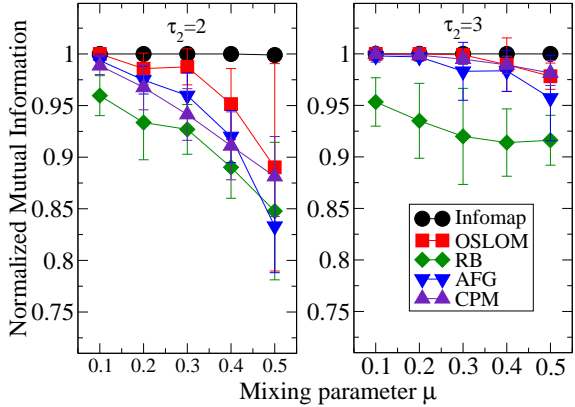}
\caption{(Color online) Comparative analysis of several
  multiresolution techniques on the LFR benchmark. The network
  parameters are now the same as for the graphs used in
  Fig.~\ref{bench1}. In particular, the network size is $10000$ and 
the cluster size spans two orders of magnitude. The two panels
correspond to $\tau_2=2$ (left) and $\tau_2=3$ (right).}
\label{fig9}
\end{center}
\end{figure}

As found in Ref.~\cite{traag11}, it is possible to find values of the
resolution parameter for RB and CPM, that make these methods
outperform both Infomap and OSLOM. This holds for AFG as well, whose
performance is essentially identical as RB. However, this is due to
the fact that the cluster sizes are too close to each other, spanning
less than one order of magnitude. This is demonstrated by Fig.~\ref{fig9},
in which we take LFR benchmark graphs with the same parameters as
those used for Fig.~\ref{bench1}. Now we have $10000$ vertices and
cluster sizes vary from $10$ to $1000$ vertices. Again, for the multiresolution methods we use the
values of the resolution parameters that give the best results. The
figure shows that the multiresolution methods fail to detect the
planted partition even for very low values of the mixing parameter
$\mu$, especially when the cluster size distribution is broader
($\tau_2=-2$). This is consistent with the results of
Figs.~\ref{bench1} and \ref{bench2}. Infomap and OSLOM, on the other hand, have a clearly
better performance, despite the fact that they do not have a tunable
resolution parameter. In particular, Infomap always detects the right
partition, for the range of $\mu$ explored here.
Most networks of current interest have many more
than $10000$ vertices, and accordingly community sizes span 
much broader ranges of values. Fig.~\ref{fig9} suggests that in such
cases the performance of multiresolution methods might become far worse.

\section{Conclusions}
\label{sec4}

We have shown that multiresolution modularity maximization
is characterized by two concurrent biases:
the tendency to merge small clusters and to split large ones. We have
seen that it is usually very difficult, and often impossible, to tune the
resolution such to avoid both biases simultaneously. Tests on
artificial benchmark graphs with community structure indeed show that
a considerable fraction of vertices is misclassified, for any value of
the resolution parameter, even when clusters are well separated and
easily identified by other methods. Since, in practical applications, one knows
very little about the community structure of the graphs at study, it
is impossible a priori to quantify the
systematic error induced by the use of modularity. Moreover, it is
not easy to implement a way to ``heal'' the partition
delivered by modularity, just because there are two sources of
errors. If modularity simply combined smaller clusters in larger ones,
as people have been thinking until now, one could hope to recover
the real partition by looking inside the clusters delivered by
modularity. Instead, since clusters can be both split and merged, the
real partition must be recovered by splitting some clusters and
merging others, and it is very difficult to
understand which clusters contain smaller ones and which others are
parts of larger clusters instead. This would require a careful
exploration of groups of clusters.

Our results hold for various types of quality functions, including
the recently introduced Constant Potts Model by Traag et
al.~\cite{traag11}. 
One could argue that, after all, multiresolution methods have a
remarkable performance in some cases (see Fig.~\ref{fig8}) and a poor
one in others (see Fig.~\ref{fig9}), just like any method, including
Infomap and OSLOM (from the same figures). This objection is however
not sustainable, since we believe that, when clusters are so weakly
connected to each other that one could even distinguish them by visual
inspection, a good method cannot fail to detect them. While this is a
shared view among scholars, it is still
unclear where to set the limit of fuzziness between subgraphs that
separates a regime in which they are clusters from one in which they
are not. This problem has attracted some attention lately~\cite{bianconi09,lancichinetti10}.
So, in the tests we reported (Figs.~\ref{fig8} and \ref{fig9}) it is
not clear up to which value of the mixing parameter $\mu$ the
subgraphs of the benchmark graphs are still ``significant'' clusters,
beyond random fluctuations. But there is no doubt that they are
cluster for very small values of the mixing parameter $\mu$.

We want to stress here that we are not advocating the superiority of
some methods over others. The problems that we point out in this paper
are probably common to many other methods. Infomap itself, for
instance, is a method based on the optimization of a global measure,
like modularity, and is likely to have a resolution limit as well,
although it probably emerges
only on large networks. In addition, it may also break random
subgraphs, although its performance is perfect for well separated
communities in all tests we have performed. OSLOM
could be also improved, since it occasionally fails to detect the
right partition for small $\mu$. Still, at variance with
multiresolution methods, neither Infomap
nor OSLOM have a tunable resolution parameter, so their
performance is quite remarkable.

We conjecture that the tendency to simultaneously merge and
split clusters is an inevitable feature of methods based on global
optimization, and that it could be more easily circumvented by local
approaches. Global optimization techniques work well when
clusters are approximately of the same size; if clusters span a broad
range of sizes, which is likely to happen on very large networks, such techniques
get confused and may fail to detect some of the clusters, even when
they are clearly identifiable. Resolution
parameters improve things, but they do not (cannot?) solve the problem.

We hope that the scientific community working on the problem of
community detection will address this issue in the future, and that general
structural limits of classes of methods will be identified and,
possibly, removed. In this way it will be possible to define safe
guidelines to design new methods that do not suffer from such problems
and that therefore could be more reliable in practical applications.

\begin{acknowledgments}
We gratefully acknowledge ICTeCollective, grant 
238597 of the European Commission.
\end{acknowledgments}

\end{document}